\documentclass[times,10pt,twocolumn]{article}
\usepackage{latex8}
\usepackage{times}
\usepackage{graphicx}
\usepackage{amsmath}
\usepackage{url}
\usepackage{subfig}

\begin{document}

\title{Phagocytes: A Holistic Defense and Protection Against Active P2P Worms}

\author{Ruichuan Chen$^\dag$, Eng Keong Lua$^\ddag$, Jon Crowcroft$^\S$, Liyong Tang$^\dag$, Zhong Chen$^\dag$\\
$^\dag$Peking University, China; $^\ddag$Carnegie Mellon University,
USA; $^\S$University of Cambridge, UK}

\maketitle
\thispagestyle{empty}

\begin{abstract}

Active Peer-to-Peer (P2P) worms present serious threats to the
global Internet by exploiting popular P2P applications to perform
rapid topological self-propagation. Active P2P worms pose more
deadly threats than normal scanning worms because they do not
exhibit easily detectable anomalies, thus many existing defenses are
no longer effective.

We propose an immunity system with \emph{Phagocytes} --- a small
subset of elected P2P hosts that are immune with high probability
and specialized in finding and ``eating'' worms in the P2P overlay.
The Phagocytes will monitor their managed P2P hosts' connection
patterns and traffic volume in an attempt to detect active P2P worm
attacks. Once detected, local isolation, alert propagation and
software patching will take place for containment. The Phagocytes
further provide the access control and filtering mechanisms for
communication establishment between the internal P2P overlay and the
external hosts. We design a novel adaptive and interaction-based
computational puzzle scheme at the Phagocytes to restrain external
worms attacking the P2P overlay, without influencing legitimate
hosts' experiences significantly. We implement a prototype system,
and evaluate its performance based on realistic massive-scale P2P
network traces. The evaluation results illustrate that our
Phagocytes are capable of achieving a total defense against active
P2P worms.

\end{abstract}

\section{Introduction}

The ability to gain control of a huge amount of Internet hosts could
be easily achieved by the exploitation of worms which self-propagate
through popular Internet applications and services. Internet worms
have already proven their capability of inflicting massive-scale
disruption and damage to the Internet infrastructure. These worms
employ normal \emph{scanning} as a strategy to find potential
vulnerable targets, i.e., they randomly select victims from the IP
address space. So far, there have been many existing schemes that
are effective in detecting such scanning worms~\cite{li08survey},
e.g., by capturing the scanning events~\cite{VirtualHpot:Provos2004}
or by passively detecting abnormal network traffic
activities~\cite{Portscant:Jung2004}.

In recent years, Peer-to-Peer (P2P) overlay applications have
experienced an explosive growth, and now dominate large fractions of
both the Internet users and traffic volume~\cite{KumarYBRR06}; thus,
a new type of worms that leverage the popular P2P overlay
applications, called \emph{P2P worms}, pose a very serious threat to
the Internet~\cite{p2p-worm-website}. Generally, the P2P worms can
be grouped into two categories: \emph{passive} P2P worms and
\emph{active} P2P worms. The passive P2P worm attack is generally
launched either by copying such worms into a few P2P hosts' shared
folders with attractive names, or by participating into the overlay
and responding to queries with the index information of worms.
Unable to identify the worm content, normal P2P hosts download these
worms unsuspectedly into their own shared folders, from which others
may download later without being aware of the threat, thus passively
contributing to the worm propagation. The passive P2P worm attack
could be mitigated by current patching systems~\cite{XieSZ08} and
reputation models~\cite{WalshS06}. In this paper, we focus on
another serious P2P worm: active P2P worm.

The active P2P worms could utilize the P2P overlay applications to
retrieve the information of a few vulnerable P2P hosts and then
infect these hosts, or as an alternative, these worms are directly
released in a hit list of P2P hosts to bootstrap the worm infection.
Since the active P2P worms have the capacity of gaining control of
the infected P2P hosts, they could perform rapid \emph{topological
self-propagation} by spreading themselves to neighboring hosts, and
in turn, spreading throughout the whole network to affect the
quality of overlay service and finally cause the overlay service to
be unusable. The P2P overlay provides an accurate way for worms to
find more vulnerable hosts easily without probing randomly selected
IP addresses (i.e., low connection failure rate). Moreover, the worm
attack traffic could easily blend into normal P2P traffic, so that
the active P2P worms will be more deadly than scanning worms. That
is, they do not exhibit easily detectable anomalies in the network
traffic as scanning worms do, so many existing defenses against
scanning worms are no longer
effective~\cite{DBLP:conf/iptps/ZhouZMICC05}.

Besides the above internal infection in the P2P overlay, the
infected P2P hosts could again mount attacks to external hosts. In
similar sense, since the P2P overlay applications are pervasive on
today's Internet, it is also attractive for malicious external hosts
to mount attacks against the P2P overlay applications and then
employ them as an ideal platform to perform future massive-scale
attacks, e.g., botnet attacks.

In this paper, we aim to develop a \emph{holistic} immunity system
to provide the mechanisms of both \emph{internal defense} and
\emph{external protection} against active P2P worms. In our system,
we elect a small subset of P2P overlay nodes, \emph{Phagocytes},
which are immune with high probability and specialized in finding
and ``eating'' active P2P worms. Each Phagocyte in the P2P overlay
is assigned to manage a group of P2P hosts. These Phagocytes monitor
their managed P2P hosts' connection patterns and traffic volume in
an attempt to detect active P2P worm attacks. Once detected, the
local isolation procedure will cut off the links of all the infected
P2P hosts. Afterwards, the responsible Phagocyte performs the
contagion-based alert propagation to spread worm alerts to the
neighboring Phagocytes, and in turn, to other Phagocytes. Here, we
adopt a threshold strategy to limit the impact area and enhance the
robustness against the malicious alert propagations generated by
infected Phagocytes. Finally, the Phagocytes help acquire the
software patches and distribute them to the managed P2P hosts. With
the above four modules, i.e., detection, local isolation, alert
propagation and software patching, our system is capable of
preventing internal active P2P worm attacks from being effectively
mounted within the P2P overlay network.

The Phagocytes also provide the access control and filtering
mechanisms for the connection establishment between the internal P2P
overlay and the external hosts. Firstly, the P2P traffic should be
contained within the P2P overlay, and we forbid any P2P traffic to
leak from the P2P overlay to external hosts. This is because such
P2P traffic is generally considered to be malicious and it is
possible that the P2P worms ride on such P2P traffic to spread to
the external hosts. Secondly, in order to prevent external worms
from attacking the P2P overlay, we hide the P2P hosts' IP addresses
with the help of scalable distributed DNS service, e.g.,
CoDoNS~\cite{1015504}. An external host who wants to gain access to
the P2P overlay has no alternative but to perform an interaction
towards the associated Phagocyte to solve an adaptive computational
puzzle; then, according to the authenticity of the puzzle solution,
the Phagocyte can determine whether to process the request.

We implement a prototype system, and evaluate its performance on a
massive-scale testbed with realistic P2P network traces. The
evaluation results validate the effectiveness and efficiency of our
proposed holistic immunity system against active P2P worms.

\textbf{Outline}. We specify the system architecture in
section~\ref{sec:SystemArchitecture}.
Sections~\ref{sec:InternalDefenses} and~\ref{sec:ExternalDefenses}
elaborate the internal defense and external protection mechanisms,
respectively. We then present the experimental design in
section~\ref{sec:ExDesign}, and discuss the evaluation results in
section~\ref{sec:ExResults}. Finally, we give an overview of related
work in section~\ref{sec:RelatedWork}, and conclude this paper in
section~\ref{Conclusions}.

\section{System Architecture}
\label{sec:SystemArchitecture}

Current P2P overlay networks can generally be grouped into two
categories~\cite{lua04survey}: \emph{structured} overlay networks,
e.g., Chord~\cite{Chord:Stoica2001}, whose network topology is
tightly controlled based on distributed hash table, and
\emph{unstructured} overlay networks, e.g.,
Gnutella~\cite{gnutella:ultrapeers02}, which merely impose loose
structure on the topology. In particular, most modern unstructured
P2P overlay networks utilize a two-tier structure to improve their
scalability: a subset of peers, called \emph{ultra-peers}, construct
an unstructured mesh while the other peers, called
\emph{leaf-peers}, connect to the ultra-peer tier for participating
into the overlay network.

\begin{figure}[tbp]
\centering
\includegraphics[angle=270,width=0.47\textwidth]{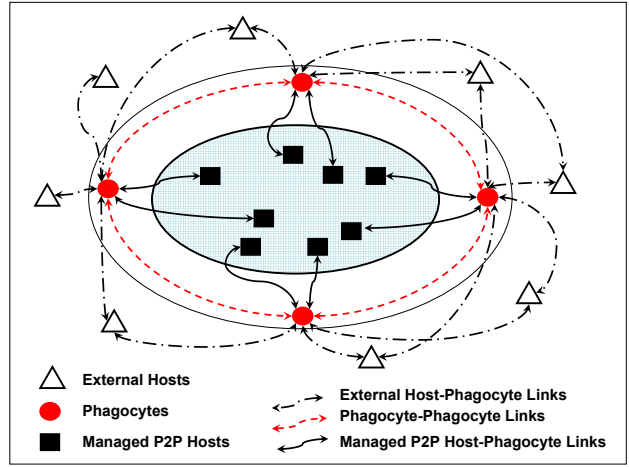}
\caption{System Architecture} \label{fig:LOVERS}
\end{figure}

As shown in Figure~\ref{fig:LOVERS}, the network architecture of our
system is similar to that of the two-tier unstructured P2P overlay
networks. In our system, a set of P2P hosts act as the Phagocytes to
perform the functions of defense and protection against active P2P
worms. These Phagocytes are elected among the participating P2P
hosts in terms of the following metrics: high bandwidth, powerful
processing resource, sufficient uptime, and applying the latest
patches (interestingly, the experimental result shown in
section~\ref{sec:ExResults} indicates that we actually do not need
to have a large percentage of Phagocytes applying the latest
patches). As existing two-tier unstructured overlay networks do, the
Phagocyte election is performed periodically; moreover, even if an
elected Phagocyte has been infected, our internal defense mechanism
(described in section~\ref{sec:InternalDefenses}) can still isolate
and patch the infected Phagocyte immediately. In particular, the
population of Phagocytes should be small as compared to the total
overlay population, otherwise the scalability and applicability are
questionable.

As a result, each elected Phagocyte covers a number of managed P2P
hosts, and each managed P2P host will belong to one closest
Phagocyte. That is, the Phagocyte acts as the proxy for its managed
hosts to participate into the P2P overlay network, and has the
control over the managed P2P hosts. Moreover, a Phagocyte further
connects to several nearby Phagocytes based on close proximity.

Our main interest is the unstructured P2P overlay networks, since
most of the existing P2P worms target the unstructured overlay
applications~\cite{p2p-worm-website}. Naturally, due to the similar
network architecture, our system can easily be deployed into the
unstructured P2P overlay networks. Moreover, for structured P2P
overlay networks, a subset of P2P hosts could be elected to perform
the functions of Phagocytes. We aim not to change the network
architecture of the structured P2P overlay networks; however, we
elect Phagocytes to form an overlay to perform the defense and
protection functions --- this overlay acts as a security wall in a
separate layer from the existing P2P overlay, thus not affecting the
original P2P operations. In the next two sections, we will elaborate
in detail our mechanisms of internal defense and external protection
against active P2P worms.

\section{Internal Defense}
\label{sec:InternalDefenses}

In this section, we first describe the active P2P worm attacks, and
then, we design our internal defense mechanism.

\subsection{Threat Model}
\label{subsec:threat_model}

Generally, active P2P worms utilize the P2P overlay to accurately
retrieve the information of a few vulnerable P2P hosts, and then
infect these hosts to bootstrap the worm infection. On one hand, a
managed P2P host clearly knows its associated Phagocyte and its
neighboring P2P hosts that are managed by the same Phagocyte; so
now, an infected managed P2P host could perform the worm infection
in several ways simultaneously. Firstly, the infected P2P host
infects its neighboring managed P2P hosts very quickly. Secondly,
the infected P2P host attempts to infect its associated Phagocyte.
Lastly, the infected managed P2P host could issue P2P key queries
with worms to infect many vulnerable P2P hosts managed by other
Phagocytes. On the other hand, a Phagocyte could be infected as
well; if so, the infected Phagocyte infects its managed P2P hosts
and then its neighboring Phagocytes. As a result, in such a
topological self-propagation way, the active P2P worms spread
through the whole system at extraordinary speed.

\subsection{Detection}
\label{subsec:detection}

Since the active P2P worms propagate based on the topological
information, and do not need to probe any random IP addresses, thus
their connection failure rate should be low; moreover, the P2P worm
attack traffic could easily blend into normal P2P traffic.
Therefore, the active P2P worms do not exhibit easily detectable
anomalies in the network traffic as normal scanning worms do.

In our system, the Phagocytes are those elected P2P hosts with the
latest patches, and they can help their managed P2P hosts detect the
existence of active P2P worms by monitoring these managed hosts'
connection transactions and traffic volume. If a managed P2P host
always sends similar queries or sets up a large number of
connections, the responsible Phagocyte deduces that this managed P2P
host is infected. Another pattern the Phagocytes will monitor is to
determine if a portion of the managed P2P hosts have some similar
behaviors such as issuing the similar queries, repeating to connect
with their neighboring hosts, uploading/downloading the same files,
etc., then they are considered to be infected.

Concretely, a managed P2P host's \emph{latest} behaviors are
processed into a \emph{behavior sequence} consisting of continuous
$\langle operation, payload \rangle$\footnote{For simplicity, we use
the abbreviation $\langle o, p \rangle$ hereafter.} pairs. Then, we
can compute the behavior similarity between any two P2P hosts by
using the \emph{Levenshtein edit distance}~\cite{Levenshtein66}.
Without loss of generality, we suppose that there are two behavior
sequences $BS_1$ and $BS_2$, in which $BS_{i=1,2} = \langle \langle
o_{i1}, p_{i1} \rangle, \cdots, \langle o_{ij}, p_{ij} \rangle,
\cdots, \langle o_{in}, p_{in} \rangle \rangle$, where $1 \leq j
\leq n$, and $n$ is the length of the behavior sequence. Further, we
can treat each behavior sequence $BS_i$ as the combination of the
\emph{operation sequence} $o_i = \langle o_{i1}, \cdots, o_{ij},
\cdots, o_{in} \rangle$ and the \emph{payload sequence} $p_i =
\langle p_{i1}, \cdots, p_{ij}, \cdots, p_{in} \rangle$. Now, we
simultaneously \emph{sort} the operation sequence $o_2$ and the
payload sequence $p_2$ of the behavior sequence $BS_2$ to make the
following similarity score $sim(BS_1, \widetilde{BS_2})$ be maximum.
To obtain the optimal solution, we could adopt the \emph{maximum
weighted bipartite matching} algorithm~\cite{graph00}; however, for
efficiency, we use the \emph{greedy} algorithm to obtain the
approximate solution as an alternative.

\begin{equation} \label{eqn:sim}
    \begin{aligned}
    \begin{cases}
        sim(BS_1, \widetilde{BS_2}) = \frac{\sum_{j=1}^n s(o_{1j}, \widetilde{o_{2j}}) \times s(p_{1j}, \widetilde{p_{2j}})}{n} \\
        s(o_{1j}, \widetilde{o_{2j}}) =
            \begin{cases}
            1 \textrm{\quad if $o_{1j} = \widetilde{o_{2j}}$} \\
            0 \textrm{\quad if $o_{1j} \neq \widetilde{o_{2j}}$}
            \end{cases} \\
        s(p_{1j}, \widetilde{p_{2j}}) = 1 - \frac{d(p_{1j}, \widetilde{p_{2j}})}{\max(length(p_{1j}), length(\widetilde{p_{2j}}))}
    \end{cases}
    \end{aligned}
\end{equation}

\noindent Here, $\widetilde{BS_2}$ denotes the sorted $BS_2$;
$\widetilde{o_{2j}}$ and $\widetilde{p_{2j}}$ denote the $j^{th}$
item of the sorted $o_2$ and ${p_2}$, respectively; $d(x, y)$ is the
Levenshtein edit distance function. Finally, we treat the maximum
$sim(BS_1, \widetilde{BS_2})$ as the similarity score of the two
behavior sequences. If the score exceeds a threshold $\theta_d$, we
consider the two P2P hosts perform similarly.

These detection operations are also performed between Phagocytes at
the Phagocyte-tier because they could be infected as well though
with latest patches. The infected Phagocytes could perform the worm
propagation rapidly; however, we have the local isolation, alert
propagation and software patching procedures in place to handle
these infected Phagocytes after detected by their neighboring
Phagocytes with the detection module as described above.

Note that, our detection mechanism is \emph{not} a substitution for
the existing worm detection mechanisms, e.g., the worm signature
matching~\cite{BrumleyNSWJ06}, but rather an effective P2P-tailored
complement to them. Specifically, some \emph{tricky} P2P worms may
present the features of mild propagation rate, polymorphism, etc.,
so they may maliciously propagate in lower speed than the aggressive
P2P worms; here, our software patching module (in
section~\ref{subsec:patching}) and several existing
schemes~\cite{1095824,WangGSZ04} can help mitigate such tricky worm
attacks.

Moreover, a few elaborate P2P worms, e.g., P2P-Worm.Win32.Hofox,
have recently been reported to be able to kill the
anti-virus/anti-worm programs on P2P hosts~\cite{p2p-worm-website};
at the system level, some local countermeasures have been devised to
protect defense tools from being eliminated, and the arms race will
continue. In this paper, we assume that P2P worms cannot disable our
detection module, and therefore, each Phagocyte can perform the
normal detection operations as expected; so can the following
modules.

\subsection{Local Isolation}

%\begin{figure}[tbp]
%\centering
%\includegraphics[angle=270,width=0.48\textwidth]{Local-Isolation}
%\caption{Local Isolation and Self-Organizing during P2P Worm
%Infection.} \label{fig:Local-Isolation}
%\end{figure}

If a Phagocyte discovers that some of its managed P2P hosts are
infected, the Phagocyte will cut off its connections with the
infected P2P hosts, and ask these infected hosts to further cut off
the links towards any other P2P hosts. Also, if a Phagocyte is
detected (by its neighboring Phagocyte) as infected, the detecting
Phagocyte immediately issues a message to ask the infected Phagocyte
to cut off the connections towards the neighboring Phagocytes, and
then to trigger the software patching module (in
section~\ref{subsec:patching}) at the infected Phagocyte; after the
software patching, these cut connections should be reestablished.
With the local isolation module, our system has the capacity of
self-organizing and self-healing. We utilize the local isolation to
limit the impact of active P2P worms as quickly as possible.

\subsection{Alert Propagation}

If a worm event has been detected, i.e., any of the managed P2P
hosts or neighboring Phagocytes are detected as infected, the
Phagocyte propagates a worm alert to all its neighboring Phagocytes.
Further, once a Phagocyte has received the worm alerts from more
than a threshold $\theta_a$ of its neighboring Phagocytes, it also
propagates the alert to all its neighboring Phagocytes that did not
send out the alert. In general, we should appropriately tune
$\theta_a$ to limit the impact area and improve the robustness
against the malicious alert propagation generated by infected
Phagocytes.

\subsection{Software Patching}
\label{subsec:patching}

The analytical study in~\cite{1103634} implied that the effective
software patching is feasible for an overlay network if combined
with schemes to bound the worm infection rate. In our system, the
security patches are published to the participating P2P hosts using
the following two procedures:

\textbf{Periodical patching:} A patch distribution service provided
by system maintainers periodically pushes the latest security
patches to all Phagocytes through the underlying P2P overlay, and
then these Phagocytes install and distribute them to all their
managed P2P hosts. Note that, we can utilize the periodical patching
to help mitigate the tricky P2P worms (in
section~\ref{subsec:detection}) which are harder to be detected.

\textbf{Urgent patching:} When a Phagocyte is alerted of a P2P worm
attack, it will immediately pull the latest patches from a system
maintainer via the direct HTTP connection (for efficiency, not via
the P2P overlay), and then install and disseminate them to all its
managed P2P hosts.

Specifically, each patch must be signed by the system
maintainer~\cite{XieSZ08}, so that each P2P host can verify the
patch according to the signature. Note that, the zero-day
vulnerabilities are not fictional, thus installing the latest
patches cannot always guarantee the worm immunity. The attackers may
utilize these vulnerabilities to perform deadly worm attacks. We can
integrate our system with some other systems, e.g.,
Shield~\cite{WangGSZ04} and Vigilante~\cite{1095824}, to defend
against such attacks, which can be found in~\cite{puzzle-Chen09}.

\subsection{Preventing Attacks on External Hosts}

As much as possible, the Phagocytes provide the containment of P2P
worms in the P2P overlay networks. Further, we utilize the
Phagocytes to implement the P2P traffic filtering mechanism which
forbids any P2P connections from the P2P overlay to external hosts
because such P2P connections are generally considered to be
malicious --- it is possible that the P2P worms ride on the P2P
traffic to spread to the external hosts. We can safely make the
assumption that P2P overlay traffic should be contained inside the
P2P overlay boundary, and any leaked P2P traffic is abnormal.
Therefore, once this leakiness is detected, the Phagocytes will
perform the former procedures for local isolation, alert propagation
and software patching.

\section{External Protection}
\label{sec:ExternalDefenses}

Our external protection mechanism aims to protect the P2P overlay
network against the external worm attacks. We hide the P2P hosts' IP
addresses to prevent external hosts from directly accessing the
internal P2P resources. This service can be provided by a scalable
distributed DNS system, e.g., CoDoNS~\cite{1015504}. Such DNS system
returns the associated Phagocyte which manages the requested P2P
host. Then, the Phagocyte is able to adopt our following proposed
computational puzzle scheme to perform the function of access
control over the requests issued by the requesting external host.

\subsection{Adaptive \& Interaction-based Puzzle}

We propose a \emph{novel} adaptive and interaction-based
computational puzzle scheme at the Phagocytes to provide the access
control over the possible external worms attacking the internal P2P
overlay.

\begin{figure}[tbp]
\centering
\includegraphics[width=0.47\textwidth]{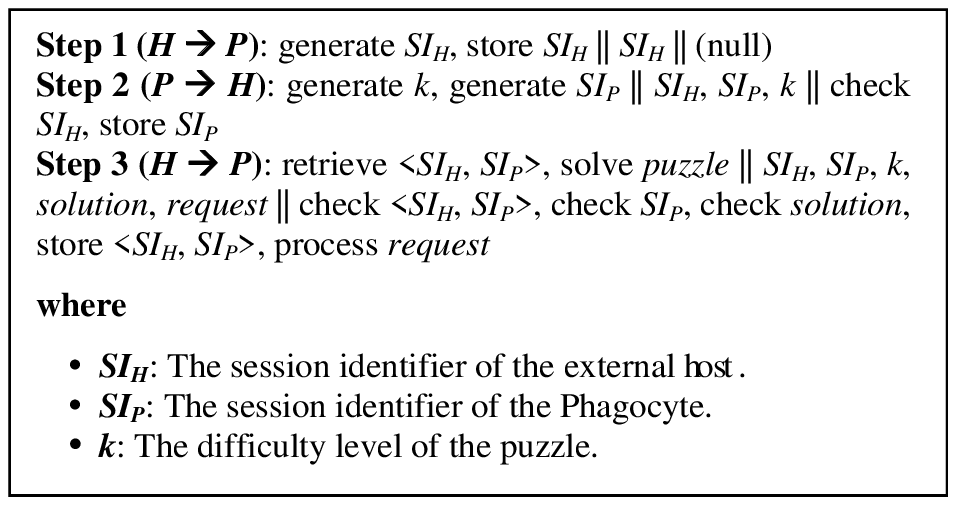}
\caption{Adaptive and Interaction-based Puzzle Scheme}
\label{fig:puzzle}
\end{figure}

Since we are interested in how messages are processed as well as
what messages are sent, for clarity and simplicity, we utilize the
annotated Alice-and-Bob specification to describe our puzzle scheme.
As shown in Figure~\ref{fig:puzzle}, to gain access to the P2P
overlay, an external host has to perform a lightweight interaction
towards the associated Phagocyte to solve an adaptive computational
puzzle; then, according to the authenticity of the puzzle solution,
the Phagocyte can determine whether to process the request.

\textbf{Step 1.} The external host $H$ first generates a $64$-bit
nonce $N_H$ as its session identifier $SI_H$. Then, the external
host stores $SI_H$ and sends it to the Phagocyte.

\textbf{Step 2.} On receiving the message consisting of $SI_H$ sent
by the host $H$, the Phagocyte $P$ adaptively adjusts the puzzle
difficulty $k$ based on the following two real-time statuses of the
network environment.

$\bullet$ \emph{Status of Phagocyte}: This status indicates the
usage of the Phagocyte's resources, i.e., the ratio of consumed
resources to total resources possessed by the Phagocyte. The more
resources a Phagocyte has consumed, the harder puzzles the Phagocyte
issues in the future.

$\bullet$ \emph{Status of external host}: In order to mount attacks
against P2P hosts effectively, malicious external hosts have no
alternative but to perform the interactions and solve many
computational puzzles. That is, the more connections an external
host tries to establish, the higher the probability that this
activity is malicious and worm-like. Hence, the more puzzles an
external host has solved in the recent period of time, the harder
puzzles the Phagocyte issues to the very external host. Note that,
since a malicious external host could simply spoof its IP address,
in order to effectively utilize the status of external host, our
computational puzzle scheme should have the capability of defending
against IP spoofing attacks, which we will describe later.

Subsequently, the Phagocyte $P$ simply generates a \emph{unique}
$64$-bit session identifier $SI_P$ for the external host according
to the host's IP address $IP_H$ (extracted from the IP header of the
received message), the host's session identifier $SI_H$ and the
puzzle difficulty $k$, as follows:

\begin{equation} \label{eqn:sip}
SI_P = HMAC_{secret} \left( IP_H | SI_H | k \right)
\end{equation}

\noindent Here, the $HMAC$ is a keyed hash function for message
authentication, and the $secret$ is a $32$-bit key which is
\emph{periodically} changed and only known to the Phagocyte itself.
Such $secret$ limits the time external hosts have for computing
puzzle solutions, and it also guarantees that an external attacker
usually does not have enough resources to pre-compute all possible
solutions in step 3.

After the above generation process, the Phagocyte replies to the
external host at $IP_H$ with the host's session identifier $SI_H$,
the Phagocyte's session identifier $SI_P$ and the puzzle difficulty
$k$. Once the external host has received this reply message, it
first checks whether the received $SI_H$ is really generated by
itself. If the received $SI_H$ is bogus, the external host simply
drops the message; otherwise, the host stores the Phagocyte's
session identifier $SI_P$ immediately. Such reply and checking
operations can effectively defend against IP spoofing attacks.

\textbf{Step 3.} The external host $H$ retrieves the $\langle SI_H,
SI_P \rangle$ pair as the global session identifier, and then tries
to solve the puzzle according to the equation below:

\begin{equation} \label{eqn:compute}
h \left( SI_H, SI_P, X \right) = Y^{ \left( k \right) }
\end{equation}

\noindent Here, the $h$ is a cryptographic hash function, the $Y^{
\left( k \right) }$ is a hash value with the first $k$ bits being
equal to $0$, and the $X$ is the puzzle \emph{solution}. Due to the
features of hash function, the external host has no way to figure
out the solution other than brute-force searching the solution space
until a solution is found, even with the help of many other solved
puzzles. The cost of solving the puzzle depends exponentially on the
difficulty $k$, which can be effortlessly adjusted by the Phagocyte.

After the brute-force computation, the external host sends the
Phagocyte a message including the global session identifier (i.e.,
the $\langle SI_H, SI_P \rangle$ pair), the puzzle difficulty, the
puzzle solution and the actual \emph{request}. Once the Phagocyte
has received this message, it performs the following operations in
turn:

\textbf{\emph{a})} Check whether the session identifier $\langle
SI_H, SI_P \rangle$ is really fresh based on the database of the
past global session identifiers. This operation can effectively
defend against replay attacks.

\textbf{\emph{b})} Check whether the Phagocyte's session identifier
$SI_P$ can be correctly generated according to equation
\ref{eqn:sip}. Specifically, this operation can additionally check
whether the difficulty level $k$ reported by the external host is
the original $k$ determined by the Phagocyte.

\textbf{\emph{c})} Check whether the puzzle solution is correct
according to equation \ref{eqn:compute}, which will also not incur
significant overhead on the Phagocyte.

\textbf{\emph{d})} Store the global session identifier $\langle
SI_H, SI_P \rangle$, and act as the overlay proxy to transmit the
request submitted by the external host. Note that, in our scheme,
the Phagocyte stores the session-specific data and processes the
actual request only after it has verified the external host's puzzle
solution. That is, the Phagocyte does not commit its resources until
the external host has demonstrated the sincerity.

Specifically in the above sequence of operations, if one operation
succeeds, the Phagocyte continues to perform the next; otherwise,
the Phagocyte cancels all the following operations, and the entire
interaction ends. More details about the puzzle design rationale can
be found in~\cite{puzzle-Chen09}.

\subsection{Comparison and Analysis}

So far, several computational puzzle
schemes~\cite{DworkN92,JuelsB99,Merkle78} have been proposed.
However, most of them consider only the status of resource
providers, so they cannot reflect the network environment
completely. Recently, an ingenious puzzle scheme,
Portcullis~\cite{ParnoWSPMH07}, was proposed. In Portcullis, since a
resource provider gives priority to requests containing puzzles with
higher difficulty levels, to gain access to the requested resources,
each resource requester, no matter legitimate or malicious, has to
compete with each other and solve hard puzzles under attacks. This
may influence legitimate requesters' experiences significantly.

Compared with existing puzzle schemes, our adaptive and
interaction-based computational puzzle scheme satisfies the
fundamental properties of a good puzzle scheme~\cite{JuelsB99}. It
treats each external host \emph{distinctively} by performing a
lightweight interaction to flexibly adjust the puzzle difficulty
according to the real-time statuses of the network environment. This
guarantees that our computational puzzle scheme does not influence
legitimate external hosts' experiences significantly, and it also
prevents a malicious external host from attacking P2P overlay
without investing unbearable resources.

In real-world networks, hosts' computation capabilities vary a lot,
e.g., the time to solve a puzzle would be much different between a
host with multiple fast CPUs and a host with just one slow CPU. To
decrease the computational disparity, some other kinds of puzzles,
e.g., memory-bound puzzle~\cite{AbadiBMW05}, could be complementary
to our scheme. Note that, with low probability, a Phagocyte may also
be compromised by external worm attackers, then they could perform
the topological worm propagation; here, our proposed internal
defense mechanism could be employed to defend against such attacks.

%In particular, to enhance the effectiveness and efficiency of our
%external protection mechanism, each Phagocyte could maintain a
%periodically refreshed blacklist of the detected malicious external
%Internet hosts, and all the Phagocytes may share their local
%blacklists with others, thus resulting in a global blacklisting
%technique which is beyond the scope of this paper.

\section{Experimental Design}
\label{sec:ExDesign}

In our experiments, we first implement a prototype system, and then
construct a massive-scale testbed to verify the properties of our
prototype system.

\begin{table*}[tbp]
    \centering
    \caption{Network Traces of Gnutella}
    \label{tab:trace}
    \footnotesize{
    \begin{tabular}{|c|c|c|c|c|c|c|}
        \hline
                                                       & Trace 1    & Trace 2     & Trace 3    & Trace 4    & Trace 5    & Trace 6   \\
        \hline
        Number (abbr., \#) of Phagocytes (Ultra-peers) & $158,985$  & $209,723$   & $51,400$   & $51,400$   & $51,400$   & $14,705$  \\
        \hline
        \# of managed P2P hosts (Leaf-peers)           & $717,025$  & $1,026,231$ & $512,448$  & $342,757$  & $257,080$  & $73,539$  \\
        \hline
        \# of Phagocytes : \# of managed P2P hosts     & $22.17\%$  & $20.44\%$   & $10.03\%$  & $15.00\%$  & $19.99\%$  & $20.00\%$ \\
        \hline
        \# of Phagocytes : \# of all P2P hosts         & $18.15\%$  & $16.97\%$   & $9.12\%$   & $13.04\%$  & $16.66\%$  & $16.66\%$ \\
        \hline
    \end{tabular}}
\end{table*}

\subsection{Prototype System}
\label{subsec:internal_prototype}

\textbf{Internal Defense.} We implement an internal defense
prototype system including all basic modules described in
section~\ref{sec:InternalDefenses}. Here, a Phagocyte monitors each
of its connected P2P hosts' latest $100$ requests. Firstly, if more
than half of the managed P2P hosts perform similar behaviors, the
responsible Phagocyte considers that the managed zone is being
exploited by worm attackers. Secondly, if more than half of a
Phagocyte's neighboring Phagocytes perform the similar operations,
the Phagocyte considers its neighboring Phagocytes are being
exploited by worm attackers. In particular, the similarity is
measured based on the equation~\ref{eqn:sim} with a threshold
$\theta_d$ of $0.5$. Then, in the local isolation module, if a
Phagocyte has detected worm attacks, the Phagocyte will cut off the
associated links between the infection zone and the connected P2P
hosts. Afterwards, in the alert propagation module, if a Phagocyte
has detected any worm attacks, it will broadcast a worm alert to all
its neighboring Phagocytes; further, if a Phagocyte receives more
than half of its neighboring Phagocytes' worm alerts, i.e.,
$\theta_a > 0.5$, the Phagocyte will also broadcast the alert to all
its neighboring Phagocytes that did not send out the alert. Finally,
in the software patching module, the Phagocytes acquire the patches
from the closest one of the system maintainers (i.e., $100$ online
trusted Phagocytes in our testbed), and then distribute them to all
their managed P2P hosts. We have not yet integrated the signature
scheme into the software patching module of our prototype system.

Note that, in the above, we simply set the parameters used in our
prototype system, and in real-world systems, the system designers
should appropriately tune these parameters according to their
specific requirements.

\textbf{External Protection.} We utilize our adaptive and
interaction-based computational puzzle module to develop the
external protection prototype system. In this prototype system, we
use SHA1 as the cryptographic hash function. Generally, solving a
puzzle with difficulty level $k$ will force an external host to
perform $2^{k-1}$ SHA1 computations on average. In particular, the
difficulty level $k$ varies between $0$ and $26$ in our system
--- this will cost an external host $0.0$ second ($k = 0$)
to $24.728$ seconds ($k = 26$) on our POWER5 CPUs. In addition, the
change cycle of the puzzle-related parameters is set to $5$ minutes.
Yet, we have not integrated our prototype system with the scalable
distributed DNS system, and this work will be part of our future
work.

\subsection{Testbed Construction}

We use the realistic network traces crawled from a million-node
Gnutella network by the Cruiser~\cite{Characterizing:Stutzbach2005}
crawler. The dedicated massive scale Gnutella network is composed of
two tiers including the ultra-peer tier and leaf-peer tier. For
historical reasons, the ultra-peer tier consists of not only modern
ultra-peers but also some \emph{legacy-peers} that reside in the
ultra-peer tier but cannot accept any leaf-peers. Specifically, in
our experiments, the ultra-peers excluding legacy-peers perform the
functions of Phagocytes, and the leaf-peers act as the managed P2P
hosts.

Then, we adopt the widely accepted GT-ITM~\cite{zegura96how} to
generate the transit-stub model consisting of $10,047$ routers for
the underlying hierarchical Internet topology. There are $10$
transit domains at the top level with an average of $10$ routers in
each, and a link between each pair of these transit routers has a
probability of $0.5$. Each transit router has an average of $10$
stub domains attached, and each stub has an average of $10$ routers,
with the link between each pair of stub routers having a probability
of $0.1$. There are two million end-hosts uniformly assigned to
routers in the core by local area network (LAN) links. The delay of
each LAN link is set to $5$ms and the average delay of core links is
$40$ms.

Now, the crawled Gnutella networks can model the realistic P2P
overlay, and the generated GT-ITM network can model the underlying
Internet topology; thus, we deploy the crawled Gnutella networks
upon the underlying Internet topology to simulate the realistic P2P
network environment. We do not model queuing delay, packet losses
and any cross network traffic because modeling such parameters would
prevent the massive-scale network simulation.

As shown in Table~\ref{tab:trace}, we list various Gnutella traces
that we use in our experiments --- with different node populations
and/or different percentages of Phagocytes.

$\bullet$ \emph{Trace} 1: Crawled by Cruiser on Sep. 27th, 2004.

$\bullet$ \emph{Trace} 2: Crawled by Cruiser on Feb. 2nd, 2005.

$\bullet$ \emph{Trace} 3: Based on trace 1, we remove a part of
Phagocytes randomly; then, we remove the \emph{isolated} Phagocytes,
i.e., these Phagocytes do not connect to any other Phagocytes;
finally, we further remove the isolated managed P2P hosts, i.e.,
these managed P2P hosts do not connect to any Phagocytes.

$\bullet$ \emph{Trace} 4: Based on trace 3, we remove a part of
managed P2P hosts randomly.

$\bullet$ \emph{Trace} 5: Based on trace 4, we further remove a part
of managed P2P hosts randomly.

$\bullet$ \emph{Trace} 6: Based on trace 1, we use the same method
as described in the generation process of trace 3. In addition, we
remove an extra part of managed P2P hosts.

\section{Evaluation Results}
\label{sec:ExResults}

\subsection{Performance Metrics}

In our experiments, we characterize the performance under various
different circumstances by using three metrics:

$\bullet$ \emph{Peak infection percentage of all P2P hosts}: The
ratio of the maximum number of infected P2P hosts to the total
number of P2P hosts. This metric indicates whether Phagocytes can
effectively defend against internal attacks.

$\bullet$ \emph{Blowup factor of latency}: This factor is the
latency penalty between the external hosts and the P2P overlay via
the Phagocytes and direct routing. This indicates the efficiency of
our Phagocytes to filter the requests from external hosts to the P2P
overlay.

$\bullet$ \emph{Percentage of successful external attacks}: The
ratio of the number of successful external attacks to the total
number of external attacks. This metric indicates the effectiveness
of our Phagocytes to prevent external hosts from attacking the P2P
overlay.

\subsection{Internal Defense Evaluation}

\begin{table*}[tbp]
    \centering
    \caption{Experimental Parameters (Internal Defense)}
    \label{tab:ex_in}
    \footnotesize{
    \begin{tabular}{|c|c|c|c|c|c|c|}
        \hline
                                                            & Experiment 1          & Experiment 2          & Experiment 3          & Experiment 4          \\
        \hline
        Initial percentage (abbr., \%) of immune Phagocytes & $100\%$ to $50\%$     & $95\%$                & $95\%$                & $95\%$                \\
        \hline
        Initial \% of immune managed P2P hosts              & $10\%$                & $0\%$ to $30\%$       & $10\%$                & $10\%$                \\
        \hline
        Initial infection \% of all vulnerable P2P hosts    & $10^{-3}\%$ to $50\%$ & $10^{-3}\%$ to $50\%$ & $10^{-3}\%$ to $50\%$ & $10^{-3}\%$ to $50\%$ \\
        \hline
        Used traces                                         & 1                     & 1                     & 1, 2, 5, 6            & 3, 4, 5               \\
        \hline
    \end{tabular}}
\end{table*}

\begin{figure*}[tbp]
    \centering
    \begin{minipage}[t]{0.243\textwidth}
        \centering
        \includegraphics[width=\textwidth]{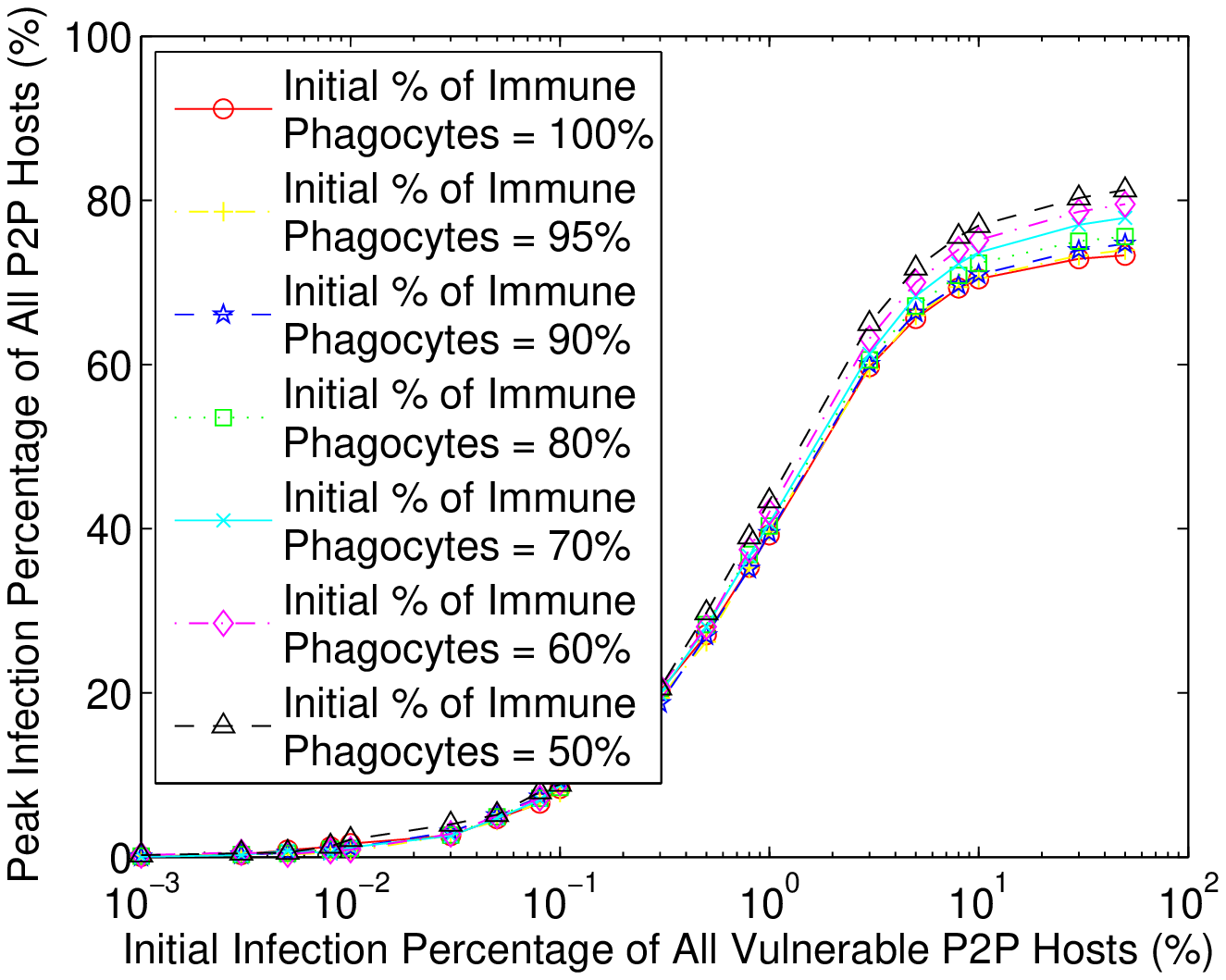}
        \caption{Impact of Immune Phagocytes}
        \label{fig:ex1}
    \end{minipage} \
    \begin{minipage}[t]{0.243\textwidth}
        \centering
        \includegraphics[width=\textwidth]{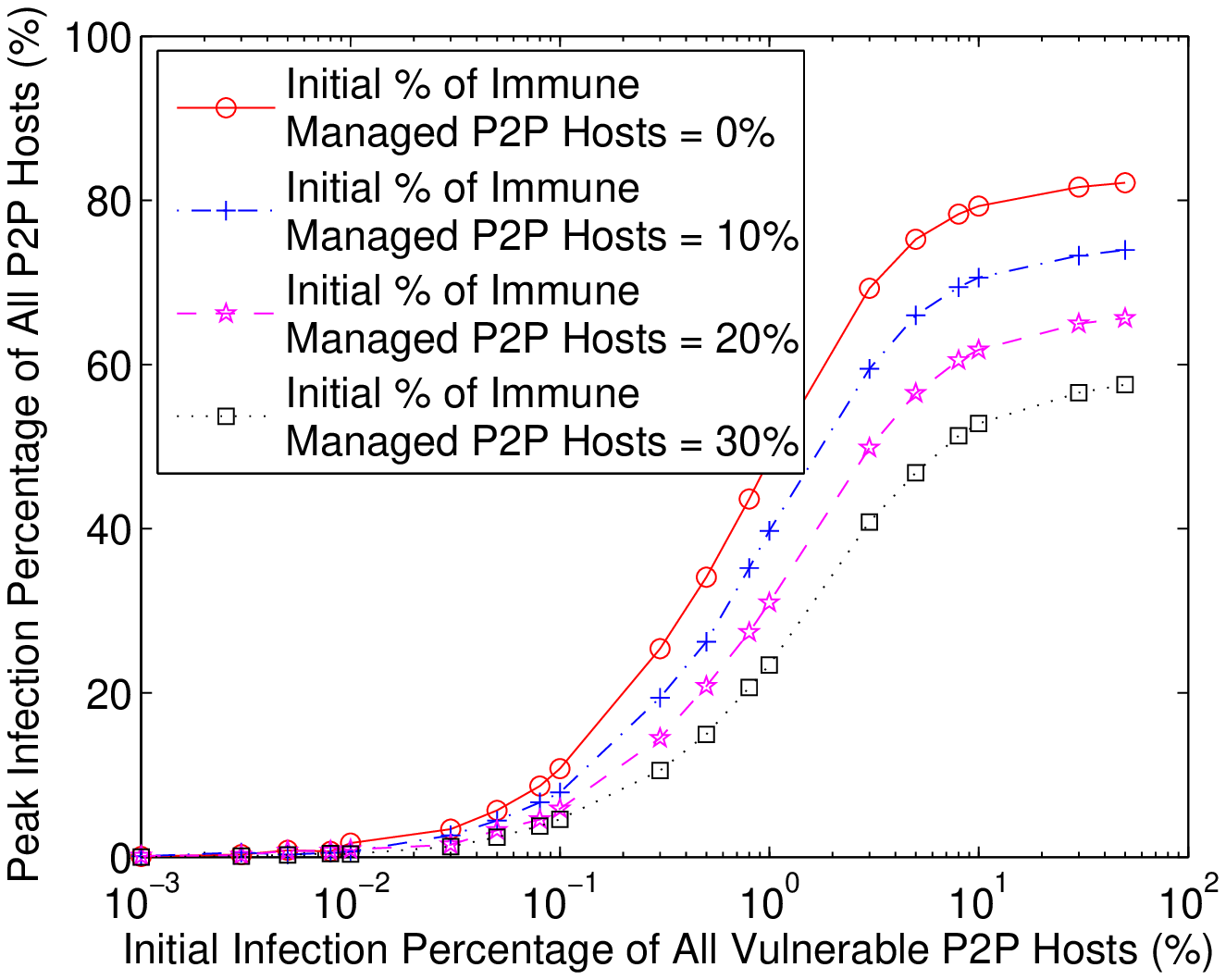}
        \caption{Impact of Immune Managed P2P Hosts}
        \label{fig:ex2}
    \end{minipage} \
    \begin{minipage}[t]{0.243\textwidth}
        \centering
        \includegraphics[width=\textwidth]{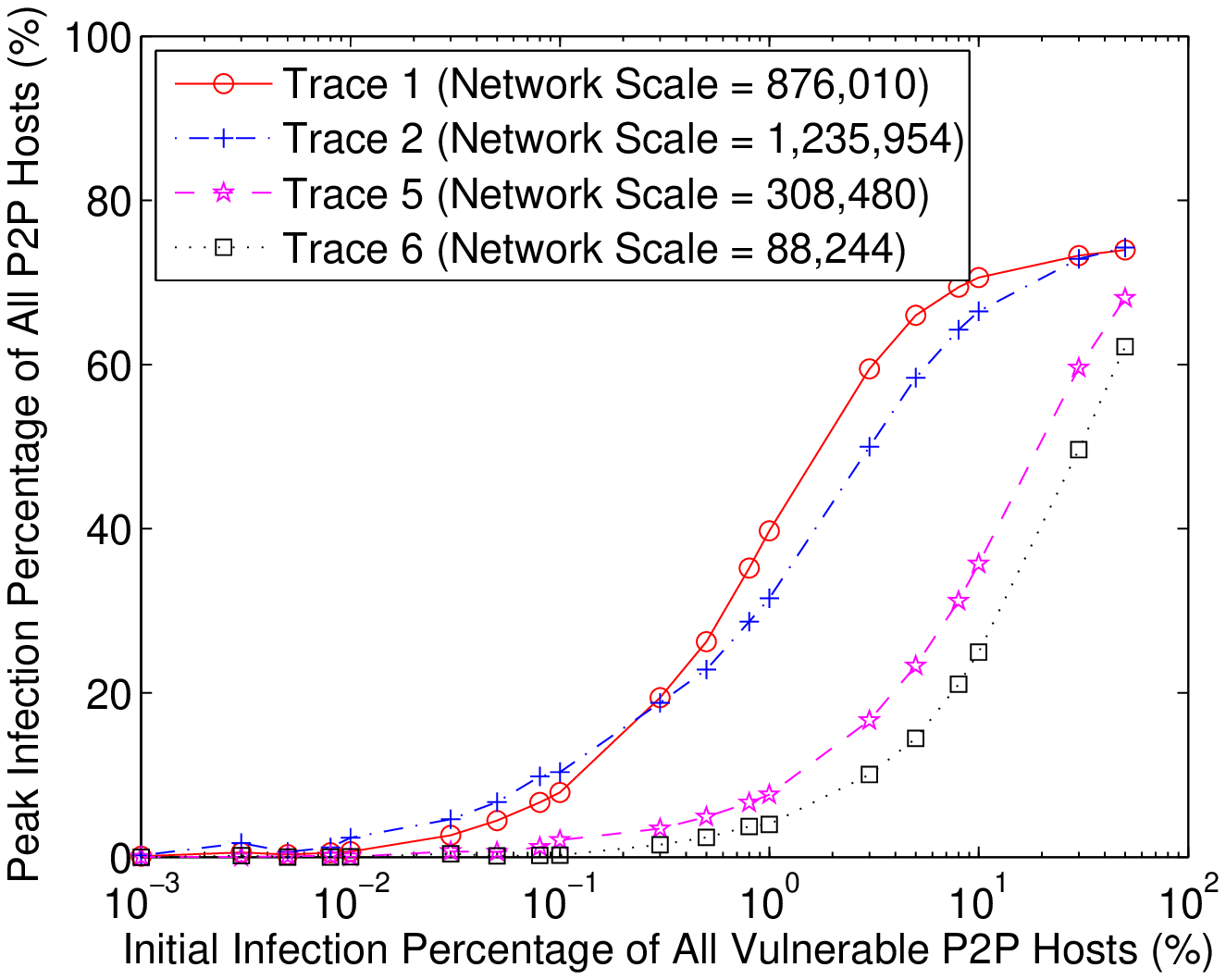}
        \caption{Impact of Network Scale}
        \label{fig:ex3}
    \end{minipage} \
    \begin{minipage}[t]{0.243\textwidth}
        \centering
        \includegraphics[width=\textwidth]{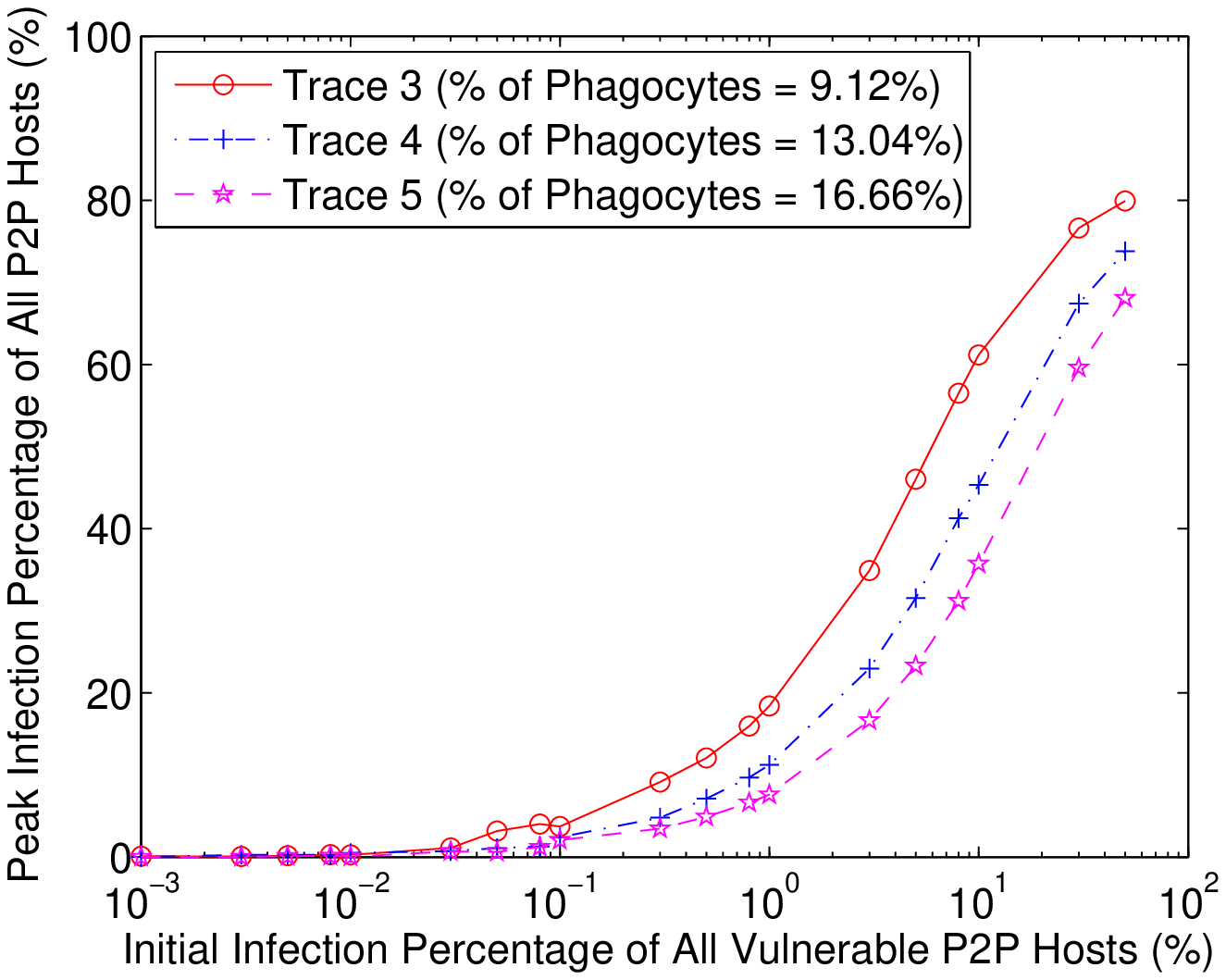}
        \caption{Impact of the Percentage of Phagocytes}
        \label{fig:ex4}
    \end{minipage}
\end{figure*}

In our prototype system, we model a percentage of Phagocytes and
managed P2P hosts being initially \emph{immune}, respectively;
except these immune P2P hosts, the other hosts are
\emph{vulnerable}. Moreover, there are a percentage of P2P hosts
being initially \emph{infected}, which are distributed among these
vulnerable Phagocytes and vulnerable managed P2P hosts uniformly at
random. All the infected P2P hosts perform the active P2P worm
attacks (described in section~\ref{subsec:threat_model}), and
meanwhile, our internal defense modules deployed at each participant
try to defeat such attacks. With different experimental parameters
described in Table~\ref{tab:ex_in}, we conduct four different
experiments to evaluate the internal defense mechanism.

\begin{figure*}[tbp]
    \centering
    \begin{minipage}[t]{0.3\textwidth}
        \centering
        \includegraphics[width=0.81\textwidth]{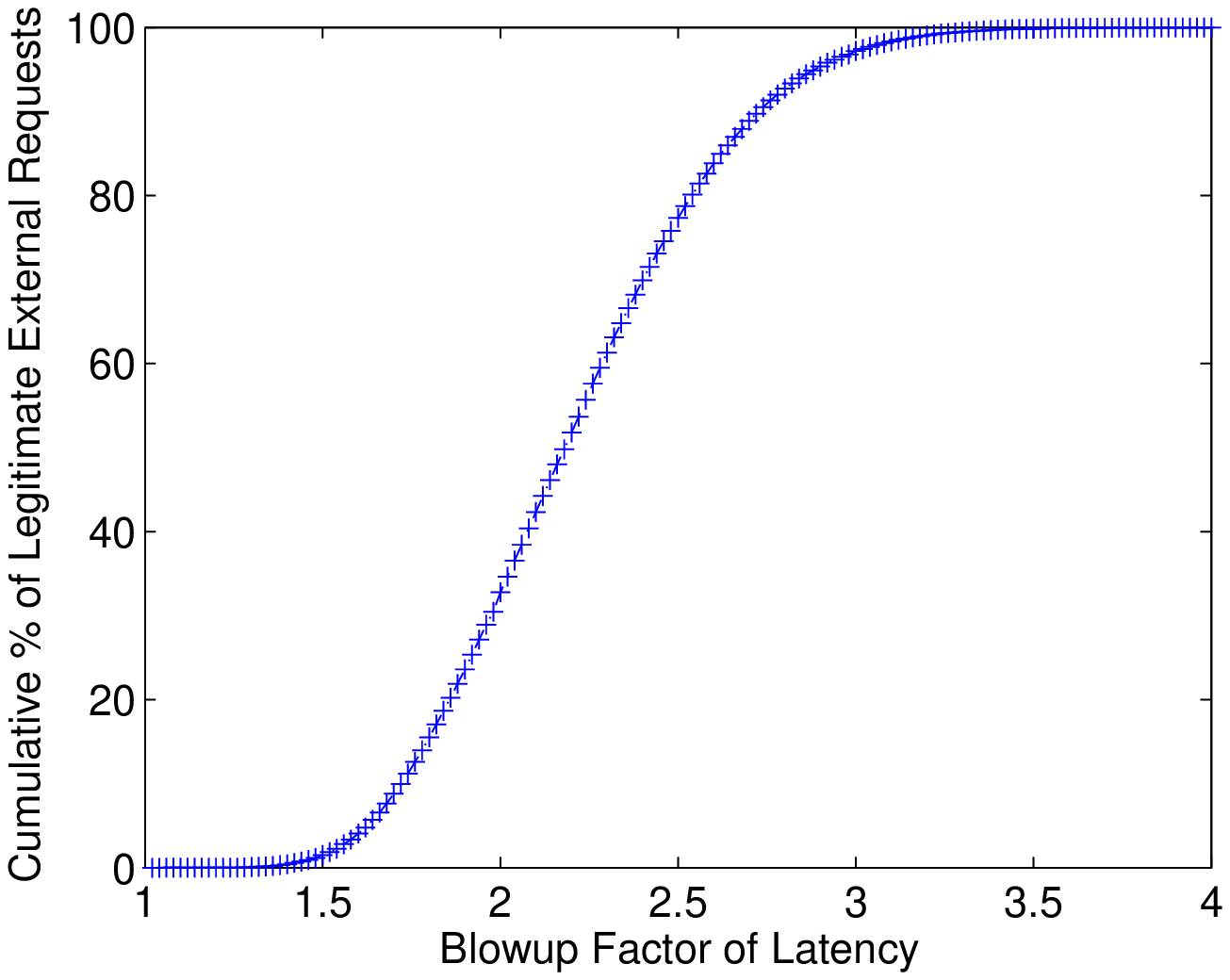}
        \caption{Cumulative distribution of the latency penalty between external hosts and P2P overlay via the Phagocytes and direct routing.}
        \label{fig:ex5}
    \end{minipage} \quad
    \begin{minipage}[t]{0.3\textwidth}
        \centering
        \includegraphics[width=0.81\textwidth]{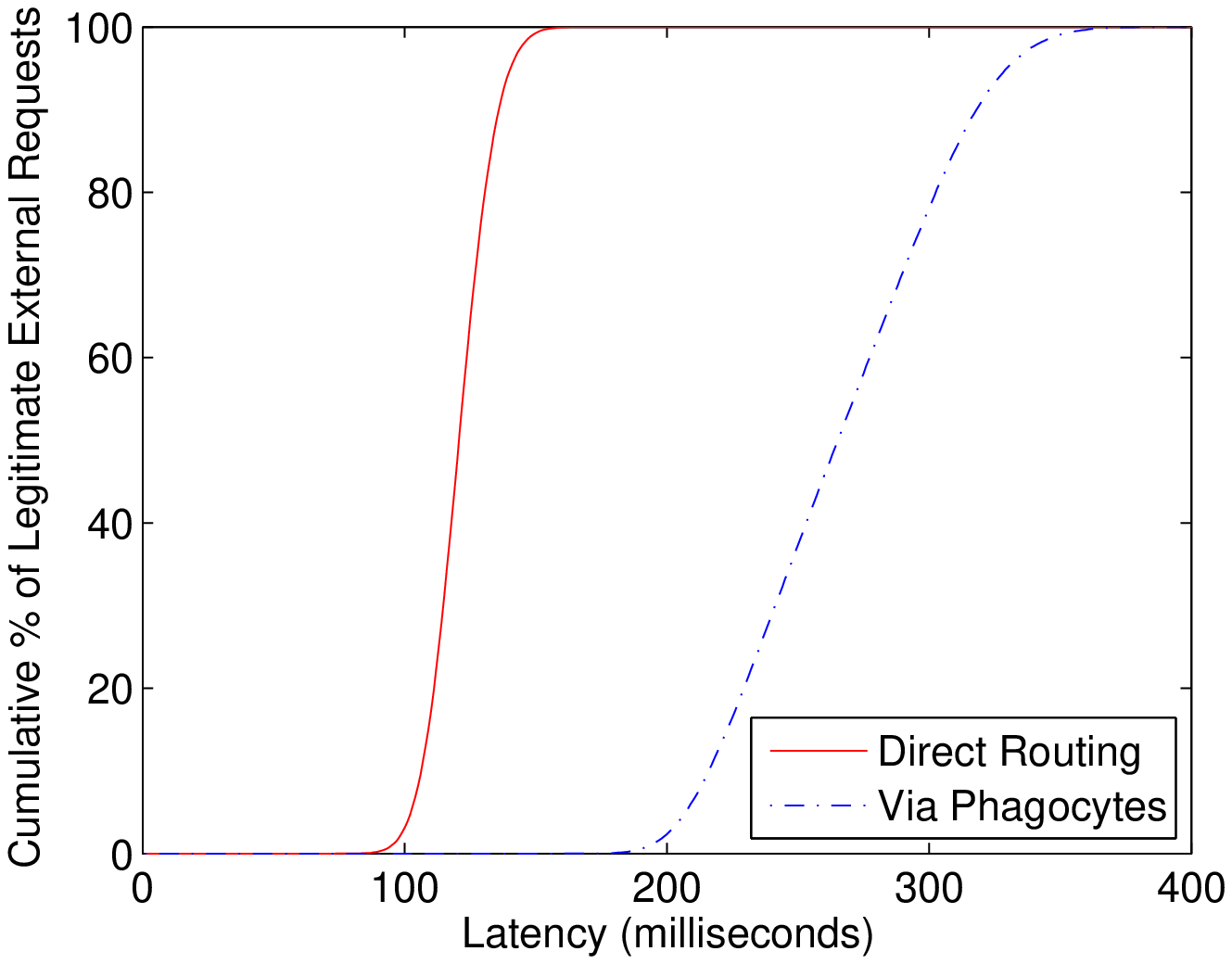}
        \caption{Cumulative distribution of the absolute latency difference between external hosts and P2P overlay via the Phagocytes and direct routing.}
        \label{fig:ex5a}
    \end{minipage} \quad
    \begin{minipage}[t]{0.3\textwidth}
        \centering
        \includegraphics[width=0.81\textwidth]{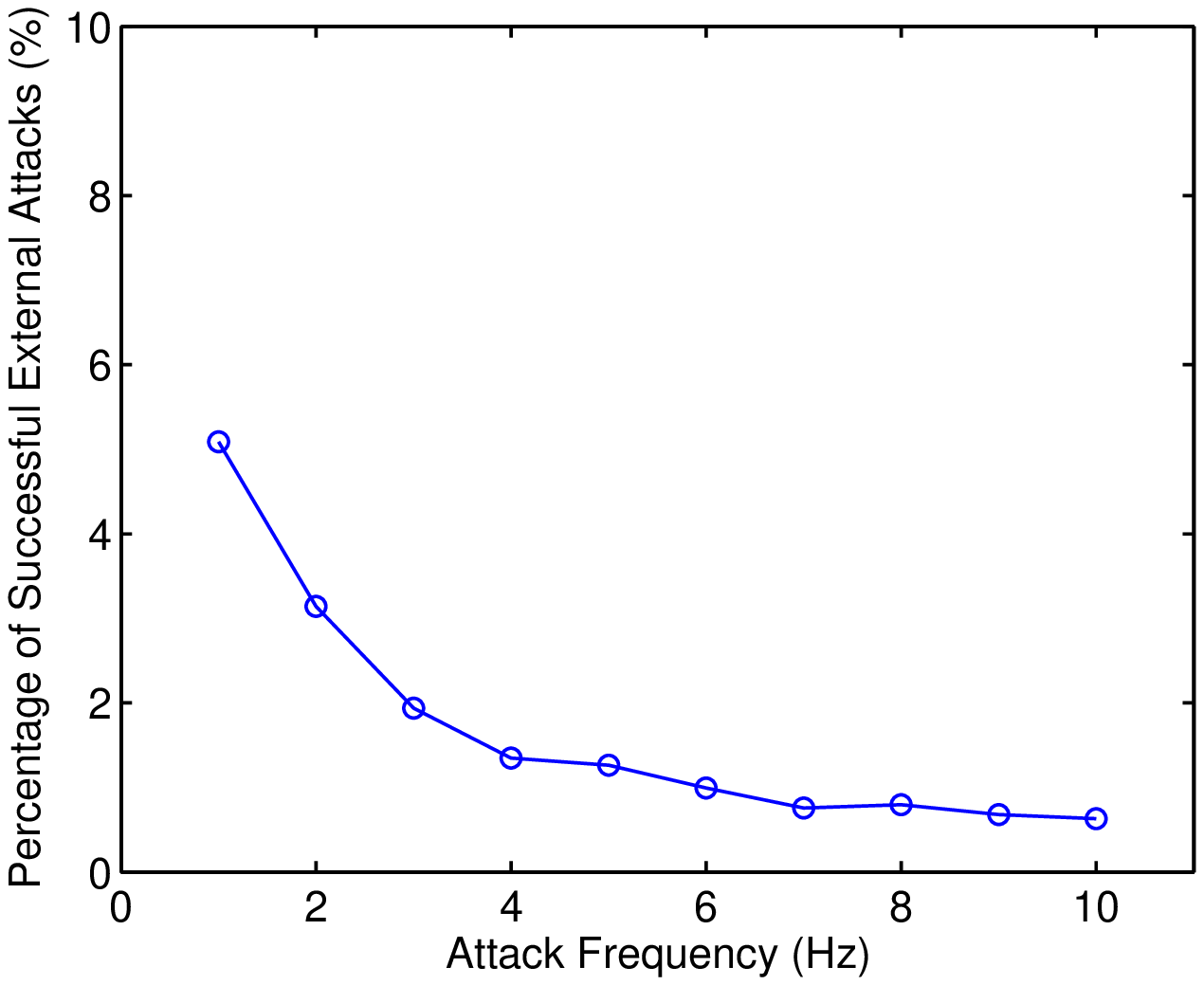}
        \caption{Effectiveness of protection against external worms attacking the P2P overlay.}
        \label{fig:ex6}
    \end{minipage}
\end{figure*}

\textbf{Experiment 1 --- Impact of immune Phagocytes:} With seven
different initial percentages of immune Phagocytes, we fix the
initial percentage of immune managed P2P hosts to $10\%$, and vary
the number of initial infected P2P hosts so that these infected
hosts make up between $10^{-3}\%$ and $50\%$ of all the vulnerable
P2P hosts. Now, we can investigate the impact of immune Phagocytes
by calculating the peak infection percentage of all P2P hosts. The
experimental result shown in Figure~\ref{fig:ex1} demonstrates that
when the initial infection percentage of all vulnerable P2P hosts is
low (e.g., $< 1\%$), the Phagocytes can provide a good containment
of active P2P worms; otherwise, the worm propagation is very fast,
but the Phagocytes could still provide the sufficient containment
--- this property is also held in the following experiments.
Interestingly, the initial percentage of immune Phagocytes does not
influence the performance of our system significantly, i.e., the
percentage of Phagocytes being initially immune has no obvious
effect. This is a good property because we do not actually need to
have high initial percentage of immune Phagocytes. Also, this
phenomenon implies that increasing the number of immune Phagocytes
does not further provide much significant defense. Thus, we can
clearly conclude that the Phagocytes are effective and scalable in
performing detection, local isolation, alert propagation and
software patching.

\textbf{Experiment 2 --- Impact of immune managed P2P hosts:} In
this experiment, for $95\%$ of Phagocytes being initially immune, we
investigate the performance of our system with various initial
percentages of immune managed P2P hosts in steps of $10\%$. The
result shown in Figure~\ref{fig:ex2} is within our expectation. The
peak infection percentage of all P2P hosts decreases with the growth
of the initial percentage of immune managed P2P hosts. Actually, in
real-world overlay networks, even a powerful attacker could
initially control tens of thousands of overlay hosts ($1\%$ -- $5\%$
in the X-axis); hence, we conclude that our Phagocytes have the
capacity of defending against active P2P worms effectively even in a
highly malicious environment.

\textbf{Experiment 3 --- Impact of network scale:}
Figure~\ref{fig:ex3} plots the performance of our system in terms of
different network scales. In traces 1, 2, 5 and 6, there are
different node populations, but the ratios of the number of
Phagocytes to the number of all P2P hosts are all around $17\%$. The
experimental result indicates that our system can indeed help defend
against active P2P worms in various overlay networks with different
network scales. Furthermore, although the Phagocytes perform more
effectively in smaller overlay networks (e.g., traces 5 and 6), they
can still work quite well in massive-scale overlay networks with
million-node participants (e.g., traces 1 and 2).

\textbf{Experiment 4 --- Impact of the percentage of Phagocytes:} In
our system, the Phagocytes perform the functions of defending
against P2P worms. In this experiment, we evaluate the system
performance with different percentages of Phagocytes but the same
number of Phagocytes. The result in Figure~\ref{fig:ex4} indicates
that the higher percentage of Phagocytes, the better security
defense against active P2P worms. That is, as the percentage of
Phagocytes increases, we can persistently improve the security
capability of defending against active P2P worms in the overlay
network. Further, the experimental result also implies that we do
not need to have a large number of Phagocytes to perform the defense
functions
--- around $10\%$ of the node population functioning as Phagocytes
is sufficient for our system to provide the effective worm
containment.

\subsection{External Protection Evaluation}

In this section, we conduct two more experiments in our prototype
system to evaluate the performance of the external protection
mechanism.

\textbf{Experiment 5 --- Efficiency:} In this experiment, we show
the efficiency in terms of the latency penalty between the external
hosts and the P2P overlay via the Phagocytes and direct routing.
Based on trace 1, we have $100$ external hosts connect to every P2P
host via the Phagocytes and direct routing in turn. Then, we measure
the latencies for both cases.

Figure~\ref{fig:ex5} plots the measurement result of latency
penalty. We can see that, if routing via the Phagocytes, about
$30\%$ and $80\%$ of the connections between the external hosts and
P2P hosts have the blowup factor of latency be less than $2$ and
$2.5$, respectively. Figure~\ref{fig:ex5a} shows the corresponding
absolute latency difference, from which we can further deduce that
the average latency growth of more than half of these connections
(via the Phagocytes) is less than $150$ms. Actually, due to the
interaction required by our proposed computational puzzle scheme, we
would expect some latency penalty incurred by routing via the
Phagocytes. With the puzzle scheme, our system can protect against
external attacks effectively which we will illustrate in the next
experiment. Hence, there would be a tradeoff between the efficiency
and effectiveness.

\textbf{Experiment 6 --- Effectiveness:} In this experiment, based
on trace 1, we have $100$ external worm attackers flood all
Phagocytes in the P2P overlay. Then, we evaluate the percentage of
successful external attacks to show the effectiveness of our
protection mechanism against external hosts attacking the P2P
overlay. For other numbers of external worm attackers, we obtain the
similar experimental results.

In Figure~\ref{fig:ex6}, the X-axis is the attack frequency in terms
of the speed of external hosts mounting worm attacks to the P2P
overlay, and the Y-axis is the percentage of successful external
attacks. The result clearly illustrates the effectiveness of
Phagocytes in protecting the P2P overlay from external worm attacks.
Our adaptive and interaction-based computational puzzle module at
the Phagocytes plays an important role in contributing to this
observation. Even in an extremely malicious environment, our system
is still effective. That is, to launch worm attacks, the external
attackers have no alternative but to solve hard computational
puzzles which will incur heavy burden on these attackers. From the
Figure~\ref{fig:ex6}, we can also find that when the attack
frequency decreases, the percentage of successful external attacks
increases gradually. However, with a low attack frequency, the
attackers cannot perform practical attacks. Even if a part of
external attacks are mounted successfully, our internal defense
mechanism can mitigate them effectively.

\section{Related Work}
\label{sec:RelatedWork}

P2P worms could exploit the perversive P2P overlays to achieve fast
worm propagation, and recently, many P2P worms have already been
reported to employ real-world P2P systems as their spreading
platforms~\cite{p2p-worm-website,1639704,shin2006malware}. The very
first work in~\cite{DBLP:conf/iptps/ZhouZMICC05} highlighted the
dangers posed by P2P worms and studied the feasibility of
self-defense and containment inside the P2P overlay. Afterwards,
several studies~\cite{1146876,1639704} developed mathematical models
to understand the spreading behaviors of P2P worms, and showed that
P2P worms, especially the active P2P worms, indeed pose more deadly
threats than normal scanning worms.

Recognizing such threats, many researchers started to study the
corresponding defense mechanisms. Specifically, Yu \emph{et al.}
in~\cite{1577951} presented a region-based active immunization
defense strategy to defend against active P2P worm attacks; Freitas
\emph{et al.} in~\cite{FreitasRRFR07} utilized the diversity of
participating hosts to design a worm-resistant P2P overlay, Verme,
for containing possible P2P worms; moreover, in~\cite{1307179}, Xie
and Zhu proposed a partition-based scheme to proactively block the
possible worm spreading as well as a connected dominating set based
scheme to achieve fast patch distribution in a race with the worm,
and in~\cite{XieSZ08}, Xie \emph{et al.} further designed a P2P
patching system through file-sharing mechanisms to internally
disseminate security patches. However, existing defense mechanisms
generally focused on the internal P2P worm defense without the
consideration of external worm attacks, so that they cannot provide
a total worm protection for the P2P overlay systems.

\section{Conclusion}
\label{Conclusions}

In this paper, we have addressed the deadly threats posed by active
P2P worms which exploit the pervasive and popular P2P applications
for rapid topological worm infection. We build an immunity system
that responds to the active P2P worm infection by using
\emph{Phagocytes}. The Phagocytes are a small subset of specially
elected P2P hosts that have high immunity and can ``eat'' active P2P
worms in the P2P overlay networks. Each Phagocyte manages a group of
P2P hosts by monitoring their connection patterns and traffic
volume. If any worm events are detected, the Phagocyte will invoke
the internal defense strategies for local isolation, alert
propagation and software patching. Besides, the Phagocytes provide
the access control and filtering mechanisms for the communication
establishment between the P2P overlay and external hosts. The
Phagocytes forbid the P2P traffic to leak from the P2P overlay to
external hosts, and further adopt a novel adaptive and
interaction-based computational puzzle scheme to prevent external
hosts from attacking the P2P overlay. To sum up, our holistic
immunity system utilizes the Phagocytes to achieve both internal
defense and external protection against active P2P worms. We
implement a prototype system and validate its effectiveness and
efficiency in massive-scale P2P overlay networks with realistic P2P
network traces.

%For future work, we would like to further characterize and study the
%effects of our internal defense and external protection strategies
%--- detection, local isolation, contagion-based alert propagation,
%software patching, computational puzzle technique, etc. This will
%help us understand the various system parameters and behaviors in
%different P2P overlay networks. We plan to port our prototype system
%over the PlanetLab~\cite{planetlabhome} as a service to substantiate
%our experimental results in wide-area network environment.

%ACKNOWLEDGMENTS are optional
%\section{Acknowledgments}
%This section is optional

%-------------------------------------------------------------------------
\bibliographystyle{latex8}
\footnotesize{
\bibliography{draft}
}

\end{document}